# Turbulence-Resilient Coherent Free-Space Optical Communications using Automatic Power-Efficient Pilot-Assisted Optoelectronic Beam Mixing of Many Modes

**Runzhou Zhang[1*], Nanzhe Hu[1*], Huibin Zhou[1], Kaiheng Zou[1], Xinzhou Su[1], Yiyu Zhou[2], Haoqian Song[1], Kai Pang[1], Hao Song[1], Amir Minoofar[1], Zhe Zhao[1], Cong Liu[1], Karapet Manukyan[1], Ahmed Almaiman[1,3], Brittany Lynn[4], Robert W. Boyd[2], Moshe Tur[5], and Alan E. Willner[1]**

1. Department of Electrical and Computer Engineering, Univ. of Southern California, Los Angeles, CA 90089, USA
2. Institute of Optics, University of Rochester, Rochester, New York 14627, USA
3. King Saud University, Riyadh 11362, Saudi Arabia
4. Naval Information Warfare Center Pacific, San Diego, CA, 92152, USA
5. School of Electrical Engineering, Tel Aviv University, Ramat Aviv 69978, Israel
* These authors contributed equally to this work.
Corresponding emails: R.Z. (runzhou@usc.edu) or A.E.W. (willner@usc.edu)

## Abstract

Atmospheric turbulence generally limits free-space optical (FSO) communications, and this problem is severely exacerbated when implementing highly sensitive and spectrally efficient coherent detection. Specifically, turbulence induces power coupling from the transmitted Gaussian mode to higher-order Laguerre-Gaussian (LG) modes, resulting in a significant decrease of the power that mixes with a single-mode local oscillator (LO). Instead, we transmit a frequency-offset Gaussian pilot tone along with the data signal, such that both experience similar turbulence and modal power coupling. Subsequently, the photodetector (PD) optoelectronically mixes all corresponding pairs of the beams' modes. During mixing, a conjugate of the turbulence experienced by the pilot tone is automatically generated and compensates the turbulence experienced by the data, and nearly all orders of the same corresponding modes efficiently mix. We demonstrate a 12-Gbit/s 16-quadrature-amplitude-modulation (16-QAM) polarization-multiplexed (PolM) FSO link that exhibits resilience to emulated turbulence. Experimental results for turbulence $D/r_0 \sim 5.5$ show up to ~20 dB reduction in the mixing power loss over a conventional coherent receiver. Therefore, our approach automatically recovers nearly all the captured data power to enable high-performance coherent FSO systems.

**Introduction**

As compared to radio-like approaches, FSO communications have gained significant interest due to its ability for higher data capacity and lower probability of interception[1–3]. Typically, an amplitude-modulated fundamental Gaussian beam is transmitted and subsequently directly detected[2]. However, system performance can be substantially increased by using coherent detection, which enables higher power sensitivity and spectrally efficient data modulation[4] (e.g., quadrature-phase-shift-keying (QPSK) and various orders of QAM[5]). Coherent detection typically uses a Gaussian LO beam at the receiver to efficiently mix with the data beam in the PD[4,6].

In general, atmospheric turbulence limits direct-detection FSO links[2,7]. However, this problem is significantly exacerbated in a coherent-detection FSO system due to the coupling of power from the fundamental Gaussian mode into other spatial modes[8–10]. In coherent FSO links, the LO and data beams are typically the same fundamental LG spatial mode (*i.e., $LG_{l=0,p=0}$*, where $l$ and $p$ represent the azimuthal and radial structures, respectively)[11]. Without turbulence-induced modal coupling, the PD efficiently mixes the single-mode data and single-mode LO since they are "mode matched" in their spatial distribution[12,13]. With turbulence, however, significant power of the data beam can be coupled into higher-order LG modes and can degrade system performance by >20 dB[8,9,14]. This occurs because the data power coupled to higher-order modes does not efficiently mix with the Gaussian LO due to different LG modes being orthogonal[14,15]. This coherent-detection degradation can occur for a PD that is: (i) free-space coupled due to mode orthogonality between the higher-modal data power and the LO[14,15], and (ii) coupled to a single-mode fiber (SMF) due to the data power in the higher-order modes not being efficiently coupled into the fiber itself[9].

Various modal-coupling mitigation approaches for coherent FSO links have been demonstrated[16–19]. One technique is to use adaptive optics wherein the following steps are performed to convert the data power back into the Gaussian mode: (a) the modal-coupling distortion is measured by a wavefront sensor, and (b) a conjugate phase is calculated by digital signal processing (DSP) and subsequently applied in a feedback loop to the data beam by a wavefront corrector[16]. Another technique is to use multi-mode digital coherent combining[17–19], wherein much of the data power in higher-order modes is captured by either a multi-mode fiber[20] or an array of SMF-based apertures[19]. Subsequently, the data power from each of multiple modes is typically recovered by a separate coherent detector with its own PD, LO (perhaps shared amongst many detectors), and DSP[17–19]. The system performance depends on the number of recovered modes (*e.g.,* 4 to 45 captured modes reduced data power loss by ~6 to ~25 dB)[17–19], and the receiver's complexity tends to be proportional to the number of detected modes such that recovering N modes uses N parallel coherent detectors[17,18]. Since turbulence may induce coupling to a large number of modes, a laudable goal would be to automatically compensate such power coupling without additional data processing or feedback, as well as to do so in a single element that efficiently scales to recover all captured modes.

In this article, we experimentally demonstrate a 12-Gbit/s 16-QAM PolM coherent FSO link that is resilient to turbulence-induced LG modal power coupling. Instead of using a receiver-based LO, we transmit a Gaussian pilot beam with a frequency offset from the data beam and which experiences similar turbulence-induced LG modal coupling as the data beam. Subsequently, a single free-space-coupled PD mixes the received multi-mode data beam with the multi-mode pilot beam that serves a similar function as the LO[7,21]. During mixing, a conjugate of the turbulence-induced modal coupling is automatically generated from the LO, which then compensates the modal coupling for the data. Specifically, each data-pilot LG modal pair efficiently mixes with each other and contributes to the intermediate-frequency (IF) signal in the electrical domain. Since the data and LO experience similar turbulence-induced modal coupling, our approach can simultaneously mix and recover almost all the captured data modes using a single PD. Experimental results for turbulence strength (*i.e.*, ratio of beam size over the Fried parameter) $D/r_0 \sim 5.5$ indicate that the pilot-assisted mixing can reduce by up to ~20 dB the mixing power loss as compared to a conventional LO heterodyne detector. Near-error-free data transmission is achieved over 200 random turbulence realizations.

## Results

### Concept and principle of automatic compensation for turbulence-induced LG modal coupling in coherent FSO links using pilot-assisted optoelectronic beam mixing

Figure 1 describes the turbulence-induced LG modal coupling and subsequent coherent detection for a conventional single-PD LO heterodyne detector (Fig. 1(a)) versus our pilot-assisted coherent detector approach (Fig. 1(b)). As shown in Fig. 1(a), a fundamental Gaussian beam (*i.e.*, $LG_{0,0}(x,y)$) carrying a coherent data channel (*e.g.*, a 16-QAM data at carrier frequency $f$, denoted as $S(t,f)$) is transmitted through turbulent atmosphere. Due to a random refractive-index distribution, the turbulence effects can induce transverse, spatially dependent wavefront distortion to the propagating Gaussian beam[22]. Moreover, since such turbulence distortion induces modal power coupling, the received electrical field (denoted as $E_{\text{data}}$) of the distorted data beam can be expressed as a superposition of many LG modes, as in Eq. (1)[8,20,23]:

$$E_{\text{data}} = S(t,f) \cdot \sum_l \sum_p a_{l,p} \cdot LG_{l,p}(x,y) = S(t,f) \cdot U, \tag{1}$$

where $LG_{l,p}(x,y)$ represents the electrical field of the LG modes with an azimuthal index of $l$ and a radial index of $p$; $a_{l,p} = \iint E_{\text{data}}(x,y) \cdot LG_{l,p}^*(x,y)dxdy$ is the complex coefficient for the corresponding $LG_{l,p}$ modal component in the distorted wavefront, and the portion of optical power coupled to the $LG_{l,p}$ mode is given by $\left|a_{l,p}\right|^2$; and $U = \sum_l \sum_p a_{l,p} \cdot LG_{l,p}(x,y)$ represents the turbulence-induced LG modal coupling. Ideally, the complex weights $a_{l,p}$ for all the LG modal components tend to satisfy that $\sum_l \sum_p \left|a_{l,p}\right|^2 \cong 1$ if the receiver aperture can collect almost the entire distorted beam's profile[23].

In a conventional single-PD LO-based heterodyne coherent detector (Fig. 1(a)), the LO is typically locally generated at the receiver with an optical frequency offset $\Delta f$ and consists of only a continuous-wave (CW) Gaussian beam (denoted as $C(f - \Delta f) \cdot LG_{0,0}(x, y)$). The square-law mixing in the PD of the coherent receiver results in photocurrent as expressed in Eq. (2)[21,24]:

$$I \propto \iint \left| C(f - \Delta f) \cdot LG_{0,0}(x, y) + S(t, f) \cdot U \right|^2 dx dy \qquad (2)$$
$$= |C|^2 + |S(t)|^2 + 2\mathrm{Re}\left[ S(t, f) \cdot C^*(f - \Delta f) \cdot \iint U \cdot LG_{0,0}^*(x, y) \, dx dy \right],$$

where $*$ denotes the conjugate of an electrical field; $\mathrm{Re}[\cdot]$ is the real part of a complex element; $I$ is the generated photocurrent; $|C|^2$ and $|S(t)|^2$ correspond to the direct-current (DC) component and the signal-signal-beating-interference (SSBI) photocurrent, respectively; and the $2\mathrm{Re}[S(t, f) \cdot C^*(f - \Delta f)]$ term generates the desired signal-LO-beating (SLB) photocurrent. However, the single-Gaussian-mode LO is not likely to efficiently mix with this multiple-LG-mode data beam due to the mode mismatch between their LG spectra, as expressed in Eq. (3)[14]:

$$\iint U \cdot LG_{0,0}^*(x, y) \, dx dy = \iint \sum_l \sum_p a_{l,p} \cdot LG_{l,p}(x, y) \cdot LG_{0,0}^*(x, y) dx dy = a_{0,0}, \qquad (3)$$

where the orthogonality among the LG modes ensures that $\iint LG_{0,0}(x, y) \cdot LG_{0,0}^*(x, y) dx dy = 1$ and $\iint LG_{l,p}(x, y) \cdot LG_{0,0}^*(x, y) dx dy = 0$ given that $l \neq 0$ or $p \neq 0$. It is shown in Eq. (3) that only the portion of the transmitted optical power that remains in the $LG_{0,0}$ mode after turbulence distortion can be efficiently mixed with the LO and utilized for recovering the in-phase and quadrature (I-Q) information. Such LG modal-coupling-based loss in captured optical data power can result in severe degradation of the mixing IF power and thus the recovered data quality[15].

Figure 1(b) illustrates the concept of our approach utilizing pilot-assisted optoelectronic beam mixing to automatically compensate the turbulence-induced LG modal coupling. In addition to the data-carrying Gaussian beam, we transmit a co-axial Gaussian beam carrying a CW pilot tone with an optical frequency offset of $\Delta f$, which is chosen to avoid the SSBI (i.e., $\Delta f$ greater than the data channel's bandwidth $B$). The electrical fields of the data beam (of carrier frequency $f$), $S(t, f)$, and that of the pilot beam, $C(f - \Delta f)$, are likely to experience similar turbulence-induced distortion and LG modal coupling due to their frequency difference being orders of magnitudes smaller than their carrier frequencies[22]. Subsequently, the data and pilot beams (*i.e.*, used like an LO at the receiver) are automatically mode matched, as expressed in Eq. (4)[25]:

$$E_{\text{pilot}} = C(f - \Delta f) \cdot U = C(f - \Delta f) \cdot \sum_l \sum_p a_{l,p} \cdot LG_{l,p}(x, y), \tag{4}$$

where $E_{\text{pilot}}$ is the received electrical field of the pilot beam. In particular, a conjugate of the turbulence-induced LG coupling $U^*$ is automatically generated from the pilot to compensate the modal coupling experienced by the distorted data beam, and the generated photocurrent can be expressed in Eq. (5):

$$I \propto \iint |C(f - \Delta f) \cdot U + S(t, f) \cdot U|^2 \, dxdy$$
$$= |C|^2 + |S(t)|^2 + 2\text{Re}[S(t, f) \cdot C^*(f - \Delta f) \cdot \iint U \cdot U^* \, dxdy], \tag{5}$$

where the $S(t, f) \cdot C^*(f - \Delta f)$ term generates the desired signal-pilot-beating (SPB) photocurrent at an IF of $\Delta f$ in the electrical domain. The turbulence-induced modal coupling is (ideally) compensated in an automatic fashion, as expressed in Eq. (6):

$$\iint U \cdot U^* dxdy = \iint \sum_l \sum_p a_{l,p} \cdot LG_{l,p}(x, y) \cdot \sum_{l'} \sum_{p'} a_{l',p'}^* \cdot LG_{l',p'}^*(x, y) \, dxdy$$
$$= \sum_l \sum_p \sum_{l'} \sum_{p'} \iint a_{l,p} \cdot LG_{l,p}(x, y) \cdot a_{l',p'}^* \cdot LG_{l',p'}^*(x, y) dxdy = \sum_l \sum_p |a_{l,p}|^2 \cong 1, \tag{6}$$

where each $LG_{l,p}$ component of the data beam is efficiently mixed with the corresponding $LG_{l,p}$ component of the pilot beam. Subsequently, almost all the captured optical power carried by higher-order LG spatial modes can contribute to the IF signal in the electrical domain and can be automatically recovered using a single square-law PD. The quality of the recovered 16-QAM signal can thus exhibit resilience to the turbulence-induced LG modal coupling loss due to the efficient mixing between the data and pilot beams.

Figures 1(a) and 1(b) also show the usable portion of the PD's electrical bandwidth. A single PD is used in both the conventional LO heterodyne detector and our pilot-assisted coherent detector. In both cases, the generated IF signal (*i.e.*, $\Delta f$) in the detector is greater than the data channel's bandwidth to avoid the SSBI and filter out the SLB or SPB photocurrent to retrieve the I-Q information. Therefore, roughly half of the PD's bandwidth can be utilized for the data that will be recovered. In terms of other coherent detector approaches (e.g., using optical hybrids and multiple PDs) that enable using more of the PDs' bandwidth, please see Discussion section for more details.

Since atmospheric turbulence tends not to induce significant depolarization effects[26], our pilot-assisted approach should be compatible with PolM techniques by transmitting pilot-data-channel pairs on each of two orthogonal polarizations. We show experimental results for such a PolM system in the Results section.

Our approach relies on using a pilot sent along with the data from the transmitter, and the pilot serves to help probe the turbulence and create a conjugate of the distortion from modal coupling. In coherent optical communications, we note that pilot-assisted techniques have been demonstrated to probe a channel's signature and apply a conjugate of that signature to help mitigate various channel impairments, including cross phase modulation[27] and a light source's phase noise[28]. More specifically, it has been shown via simulation that turbulence-induced inter-channel crosstalk can be reduced by mixing a pilot Gaussian beam and data-carrying LG beams in a coherent mode-division-multiplexed FSO link[29]. In that approach, the pilot Gaussian beam acquires the turbulence signature, is converted to different LG modes at the receiver, generates a conjugate of turbulence for each corresponding data LG beams in the PD, and recovers each of the multiplexed data channels using a separate PD.

**Experimental setup of 12-Gbit/s PolM coherent FSO communications with emulated turbulence**

We experimentally demonstrate the compatibility of the pilot-assisted optoelectronic beam mixing approach with a PolM communication system. As shown in Fig. 2, we transmit a pair of data-carrying and pilot Gaussian beams on both X and Y polarizations in a coherent FSO link. A 6-Gbit/s 16-QAM data channel at a wavelength of $\lambda_1 \sim 1.55$ μm is generated, amplified by an erbium-doped fiber amplifier (EDFA), and equally split into two copies. One copy is delayed using a > 15-m SMF to decorrelate the data channels and two independent data channels are individually combined with another pilot tone at a wavelength of $\lambda_2$ (with a frequency offset of $\sim 2.6$ GHz from $\lambda_1$). The polarizations of signals and pilots are adjusted and subsequently combined by a polarization beam combiner to transmit PolM 16-QAM signals. The total optical power including the pilot and data beams is $\sim 7$ dBm at each of the polarizations. The optical signal is coupled to free space by an optical collimator (Gaussian beam size in diameter $D \sim 2.2$ mm), is distorted by a rotatable turbulence emulator (see Methods section for more details), and propagates in free space for $\sim 1$ m. In this demonstration, we emulate different strengths of atmospheric turbulence by using two separate turbulence emulators with different Fried parameters $r_0$ of 1.0 mm and 0.4 mm. The emulated turbulence distortion for the transmitted Gaussian beam is characterized by the ratio of beam size over the Fried parameter[22], which are $D/r_0 \sim 2.2$ and $D/r_0 \sim 5.5$ for the two emulators.

At the receiver, we demultiplex one polarization at a time by using a half-wave plate (HWP) cascaded with a polarizer. The receiver has an aperture diameter of $\sim 10$ mm. We measure the spatial amplitude and phase profiles of the turbulence-distorted beam and calculate its LG decomposition using the off-axis holography[30] (see Methods section for more details). After polarization demultiplexing, the distorted beam is equally split into two copies that are sent to the pilot-assisted coherent detector and a conventional single-PD LO heterodyne detector.

In the pilot-assisted beam mixing approach, the entire spatial profiles of the distorted data and pilot beams are focused into a free-space-coupled InGaAs PD (3-dB bandwidth < 3.5 GHz)[31] using an aspheric lens with a focal length and a numerical aperture (NA) of 16 mm and $\sim 0.79$, respectively. The coupling efficiency of the received Gaussian beam, defined as the ratio of received optical power detected by the PD without turbulence effects, is measured to be > 92%. The generated photocurrent is recorded by a real-time digital oscilloscope and the I-Q information of the data channel is subsequently retrieved by using off-line DSP algorithms (see Methods section for more details).

For comparison, we measure the system performance using a conventional LO heterodyne detector (the pilot $\lambda_2$ is turned off). At this receiver, we set the same IF value as the pilot-assisted receiver to provide a fair comparison. The distorted Gaussian beam is coupled to an SMF via a collimator (aperture diameter ~3.5 mm), amplified by an EDFA, and mixed with an LO (at the same wavelength $\lambda_2$ as the pilot) at an SMF-coupled PD. The received optical signal is amplified by the EDFA to meet the power sensitivity requirement of the SMF-coupled PD. The electrical signal is subsequently recorded by a real-time digital oscilloscope and processed to retrieve the data channel's I-Q information by using the same off-line DSP algorithms as the pilot-assisted detector. Note that we measure the optical power loss and electrical mixing power loss of this conventional LO heterodyne detector (to be shown in Fig. 3) without using the EDFA at this receiver. The mixing power loss is measured at the IF of ~2.6 GHz in the electrical domain.

We emulate the atmospheric turbulence effects for this ~1-m FSO link using a single rotatable phase plate. In general, the turbulence effects can be emulated with higher accuracy using multiple phase plates[22]. To address the accuracy of our turbulence emulation, we simulate the system degradation including optical and electrical mixing power loss using the single and multiple random phase screen (RPS) models separately; the simulation results in Suppl. Fig. 1 and Suppl. Fig. 2 show similar power loss distributions and trends for both 1-RPS and 5-RPS models (see Supplementary Information for more details).

**Characterization of optical power loss and electrical mixing power loss for the pilot-assisted detector under different turbulence strengths**

We measure the turbulence-induced optical power loss and electrical mixing power loss using the conventional LO and pilot-assisted detectors on both polarizations over 1000 random realizations of emulated turbulence. The measured optical and mixing power loss for both receivers are shown in Fig. 3(a) and Fig. 3(b), respectively.

As shown in Fig. 3(a), for both X and Y polarizations, the stronger turbulence induces <2 dB optical power loss for the pilot-assisted detector because the free-space-coupled PD can detect most of the captured optical power. However, the optical power loss using the conventional LO-based detector ranges from ~2 dB to ~22 dB and from ~7 dB to ~30 dB under the turbulence strengths of $D/r_0 \sim 2.2$ and $D/r_0 \sim 5.5$, respectively. This is due to that the LO-based detector can hardly capture the optical power that are coupled to non-fundamental Gaussian modes by turbulence effects.

As shown in Fig. 3(b), the pilot-assisted detector has an electrical mixing power loss of <3 dB and <6 dB over 99% weaker turbulence realizations ($D/r_0 \sim 2.2$) and 90% stronger turbulence realizations ($D/r_0 \sim 5.5$), respectively. For comparison, the conventional LO-based detector suffers from a mixing power loss of up to ~28 dB over 99% weaker turbulence realizations and 90% stronger turbulence realizations. The relatively lower mixing power loss for the pilot-assisted detector is due to that the efficient mixing of the pilot and data beams is likely to recover almost all the data power from the captured LG modes.

We further validate the measured results by 1-RPS simulation (see Supplementary Information for more details). As shown in Fig. 3(c), the simulation results indicate that the pilot-assisted detector can effectively reduce both the average optical and mixing power loss as compared to the conventional LO-based detector when the turbulence strength $D/r_0$ varies from ~1 to ~7. Moreover, the experimentally measured data points are generally in agreement with the simulation results; in particular, for the turbulence strength $D/r_0 \sim 5.5$, the pilot-assisted detector exhibits ~14 dB and ~11.6 dB higher average optical power and electrical mixing power, respectively, than the conventional LO-based detector.

## 12-Gbit/s 16-QAM PolM data transmission through emulated turbulence

We demonstrate a 12-Gbit/s PolM FSO data transmission under emulated turbulence effects, with each polarization carrying a 1.5-GHz Nyquist-shaped 16-QAM data channel. We measure the turbulence-induced LG spectra of which the azimuthal (*i.e.*, $l$) and radial (*i.e.*, $p$) indices range from -5 to +5 and from 0 to 10, respectively. The complex wavefront is measured using off-axis holography (see Methods for more details)[30]. As shown in Fig. 4(a), both the conventional single-PD LO and pilot-assisted detectors can achieve near-error-free performance without turbulence effects, recovering error vector magnitudes (EVM) ~8% for the 16-QAM data. This is due to that most of the data power is carried by the fundamental Gaussian mode, as indicated in the measured LG spectrum.

Under one example realization of the weaker turbulence ($D/r_0 \sim 2.2$), as shown in Fig. 4(b), the measured LG spectrum shows that the power of the transmitted Gaussian mode is mainly coupled to some neighboring LG modes. Such power loss on the Gaussian mode can be mitigated by using an optical amplifier for the conventional LO heterodyne receiver, resulting in EVMs not significantly degraded. However, as shown in Fig. 4(c) and Fig. 4(d), under these two example realizations of the stronger turbulence ($D/r_0 \sim 5.5$), turbulence effects can induce an optical power loss of >25 dB for the $LG_{0,0}$ mode and the measured LG spectra indicate power coupling to a large number of higher-order LG modes. The conventional LO heterodyne detector suffers from severe data quality degradation for both polarizations (EVMs ~16% in Fig. 4(c)) and it can hardly recover the 16-QAM data, EVMs >20% in Fig. 4(d). The performance of the pilot-assisted detector is not significantly affected by the stronger turbulence effects and it can recover the 16-QAM data with EVMs ranging from ~8% to ~10% for both realizations. This is because the turbulence-induced LG modal power coupling can be automatically compensated by the pilot-data mixing and almost all the captured data-carrying LG modes can be efficiently recovered.

To investigate the turbulence effects on the SLB or SPB signal of both detectors, we also measure the electrical spectra for these example turbulence realizations. As shown in Suppl. Fig. 4, compared to the case of no turbulence effects, we measure up to ~3 dB signal-to-noise ratio (SNR) degradation of the SPB for the pilot-assisted detector, while the SNR degradation of the SLB for the conventional LO-based detector can be up to ~18 dB by these turbulence realizations (see Supplementary Information for more details).

To evaluate the robustness of the pilot-assisted coherent detector against random turbulence realizations, we measure the bit-error-rate (BER) values of the 16-QAM data channels carried by both polarizations over 100 random turbulence realizations. For comparison, we conduct the same measurements for the conventional single-PD LO heterodyne detector. Note that we measure the BER performance for one polarization at a time due to limitations of our measurement setup. Therefore, the BER values with the same realization label correspond to different random turbulence realizations and cannot be directly compared. As shown in Fig. 5, under the weaker turbulence effects ($D/r_0 \sim 2.2$), both approaches can achieve similar BER performance below the 7% forward-error-correction (FEC) limit for almost all the 200 random realizations. Under the stronger turbulence effects ($D/r_0 \sim 5.5$), the BER performance of the conventional LO heterodyne detector degrades and does not achieve the 7% FEC limit for some realizations; while the pilot-assisted detector still shows resilience to the turbulence distortion and achieves BER values below the 7% FEC limit for all 200 random turbulence realizations.

To better characterize the performance of the pilot-assisted coherent detector, we also measure the BER as a function of the transmitted power; we find power penalties of ~3 dB for both polarizations under one example random realization of the stronger turbulence, as shown in Suppl. Fig. 5 (See Supplementary Information for more details).

## Discussion

In our approach, a pilot-assisted coherent detector is used to achieve a 12-Gbit/s, 1-m link that exhibits resilience to emulated turbulence, and there are issues that should be discussed to elucidate various challenges.

The following issues are related to the electrical spectrum:

*(i) Data channel bandwidth*. Our data baud rate (*i.e.*, 1.5 GHz) is limited by the PD's 3-dB bandwidth, *i.e.*, ~3.5 GHz. However, free-space-coupled PDs with a bandwidth of ~49 GHz have been reported[32], such that it might be possible to achieve >100-Gbit/s rates.

*(ii) Utilization of PD bandwidth*. In our concept and experiment, we have compared the pilot-assisted approach to a conventional-but-simple, single-PD heterodyne approach. As stated in the Results section, we state that the electrical bandwidth usable for the data channel in both cases is roughly half the PD bandwidth. However, conventional-style coherent detectors can use multiple PDs and various types of optical hybrids, such that more of the electrical bandwidth can be used for the data channel. For example, heterodyne detectors using two PDs and intradyne detectors using four PDs have the potential to utilize up to 2x or 4x the electrical bandwidth, respectively, as our demonstration[6]. Since the multiple-PD approaches typically use single-mode components, it remains an open question as to whether our pilot-assisted approach can be realized with multiple-PD approaches and optical hybrids such that more of our electrical bandwidth can be used for the data channel. In addition, we believe that we could potentially nearly double our data channel bandwidth by reducing the frequency gap between the data and pilot[21]. This frequency gap reduction might make use of some advanced I-Q recovery techniques to mitigate the SSBI, such as the iterative estimation and cancellation[33].

Below are a few mode-related performance issues:

***(i) Other modal bases***. We utilize the LG modal basis to analyze the modal coupling. However, the pilot-assisted approach could also utilize other bases (e.g., Hermite-Gaussian[17]). Importantly, our approach does not *a priori* need to specify the employed basis because our approach is "automatic" and the pilot and data can be described as combinations of different bases.

***(ii) Beam divergence and truncation***. A Gaussian beam diverges in a longer link, and the beam can be truncated by a limited-size receiver aperture[34]. Truncation can cause power loss for both the data and pilot beams, resulting in mixing power loss in the electrical domain. Moreover, truncation can cause modal coupling to higher-order modes[34], which should be automatically mixed in the PD.

***(iii) Compatibility with fiber-coupled PDs***. We utilize a free-space-coupled PD. However, it remains unclear as to whether our pilot-assisted approach can be realized for fiber-coupled PDs. The challenge is that our approach relies on many modes impinging on the PD, but fiber-coupled PDs tend to be single-mode. One technique might be to use a multi-mode-fiber-coupled PD[20].

## Methods

### Experimental emulation of atmospheric turbulence effects with Kolmogorov power spectrum statistics

We experimentally emulate the turbulence-induced distortion by utilizing glass plates[35] of which the refractive index distributions are fabricated to emulate Kolmogorov power spectrum statistics[15]. Two rotatable glass plates are separately employed in the experiment with different Fried parameters $r_0$ of 1.0 mm (weaker turbulence effects) and 0.4 mm (stronger turbulence effects). Different random turbulence realizations are implemented by rotating the single glass plate to different orientations. The diameter of the transmitted

Gaussian beam is ~2.2 mm. The data-carrying Gaussian beams are distorted by the glass plate and then propagate in free space for a distance of ~1 m before reaching the receiver. The strength of turbulence distortion is given by the ratio of beam diameter over the Fried parameter[22], *i.e.*, $D/r_0$.

In this demonstration, we utilize a single phase plate to emulate turbulence distortions for this ~ 1-m FSO link. However, a multiple-phase-plate emulation can generally provide a higher accuracy for emulating the atmospheric turbulence effects[22]. To illustrate the validity of our emulation method, we simulate 1-RPS and 5-RPS turbulence effects; find similar trends for turbulence-induced system degradations (see Supplementary Information for more details). We note that our turbulence emulation provides an approximation of the Gaussian beam's propagation in turbulent medium and may not fully reflect the effects of real atmospheric turbulence. To further enhance the accuracy of turbulence emulation, one could potentially apply some advanced modeling or emulation methods[22,36].

**Off-axis holography for complex wavefront and LG spectrum measurement**

We utilize off-axis holography to measure the complex wavefront (*i.e.*, amplitude and phase) of the distorted Gaussian beam and its corresponding LG spectrum. An off-axis reference Gaussian beam (beam diameter ~7 mm) on the same wavelength as the distorted pilot Gaussian beam is incident on the infrared camera with a tilted angle. We record the off-axis interferogram and apply digital image processing to extract the complex wavefront (See Supplementary Information for more details). The data-carrying beam is turned off when we measure the complex wavefront of the turbulence-distorted pilot beam.

After the complex wavefront of the distorted Gaussian beam is obtained, we decompose it into a two-dimensional LG modal spectrum of which the two indices $l$ and $p$ range from -5 to +5 and from 0 to 10, respectively, as expressed in Eq. (7) :

$$a_{l,p} = \iint E_{\text{rec}}(x,y) \cdot LG_{l,p}^*(x,y)dxdy, \tag{7}$$

where $E_{rec}(x,y)$ and $LG_{l,p}(x,y)$ are the measured complex field of the distorted Gaussian beam and the theoretical complex field of an $LG_{l,p}$ mode, respectively. The ratio of optical power coupling to the $LG_{l,p}$ mode is given by $\left|a_{l,p}\right|^2$.

**DSP for retrieving the I-Q information in the coherent detection**

The detected electrical signal is sampled by a real-time oscilloscope (20-GHz bandwidth and 50-GSa/s sampling rate) and recorded for off-line DSP. The recorded signals from the pilot-assisted coherent detector and the conventional single-PD LO heterodyne detector are processed by the same DSP procedures. Each signal is filtered by a root-raised-cosine finite impulse response filter with a roll-off factor of 0.1, and the filtered signal is subsequently equalized using a constant modulus algorithm (CMA). After the CMA equalization, carrier frequency offset estimation and carrier phase recovery are sequentially performed to reduce the frequency and phase difference between the signal and the LO (or pilot). Finally, the EVM and BER of the demodulated signal are calculated to evaluate the quality of data transmission. The EVM of the detected signal is calculated using the Eq. (8) as follows[5]:

$$\text{EVM} = \sqrt{\frac{1}{N \cdot \max_i |\widehat{x_i}|^2} \cdot \sum_{i=1}^N |x_i - \widehat{x_i}|^2} \times 100\%, \tag{8}$$

where the $x_i$ and $\widehat{x_i}$ represent the transmitted and recovered data symbols, respectively; $N$ is the total number of detected symbols. In this demonstration, ~180,000 symbols are collected to calculate the EVMs and BERs of 16-QAM data signals.

**Data availability:** All data, theory details, simulation details that support the findings of this study are available from the corresponding authors on reasonable request.

**Acknowledgments**

This work is supported by Office of Naval Research through a MURI award N00014-20-1-2558; Defense Security Cooperation Agency (DSCA 4441006051); Airbus Institute for Engineering Research; Vannevar Bush Faculty Fellowship sponsored by the Basic Research Office of the Assistant Secretary of Defense (ASD) for Research and Engineering (R&E) and funded by the Office of Naval Research (ONR) (N00014-16-1-2813); Qualcomm Innovation Fellowship (QIF); Naval Information Warfare Center Pacific (N6600120C4704).

**Author information**

**Affiliations:**

1. Department of Electrical and Computer Engineering, Univ. of Southern California, Los Angeles, CA 90089, USA

   Runzhou Zhang, Nanzhe Hu, Huibin Zhou, Kaiheng Zou, Xinzhou Su, Haoqian Song, Kai Pang, Hao Song, Amir Minoofar, Zhe Zhao, Cong Liu, Karapet Manukyan, Ahmed Almaiman, and Alan E. Willner

2. Institute of Optics, University of Rochester, Rochester, New York 14627, USA

   Yiyu Zhou, and Robert W. Boyd

3. King Saud University, Riyadh 11362, Saudi Arabia

   Ahmed Almaiman

4. Naval Information Warfare Center Pacific, San Diego, CA, 92152, USA

   Brittany Lynn

5. School of Electrical Engineering, Tel Aviv University, Ramat Aviv 69978, Israel

   Moshe Tur

**Author contributions:** All the authors contributed to the interpretation of the results and manuscript writing. R.Z. conceived the idea; R.Z. and N.H. conducted the experiment; H.Z., K.Z., and X.S. performed the digital signal processing; Y. Z. helped implement the off-axis holography; H.S., K.P., H.S., A.M., Z.Z., C.L., K.M., and A.A. carried out the measurement and data analysis; B.L., R.W.B, M.T., and A.E.W. provided the technical support. The project was supervised by A.E.W.

**Corresponding author:** Correspondence to Runzhou Zhang and Alan E. Willner. Corresponding emails: runzhou@usc.edu, willner@usc.edu.

**Competing Interests:** The authors declare no competing interests.

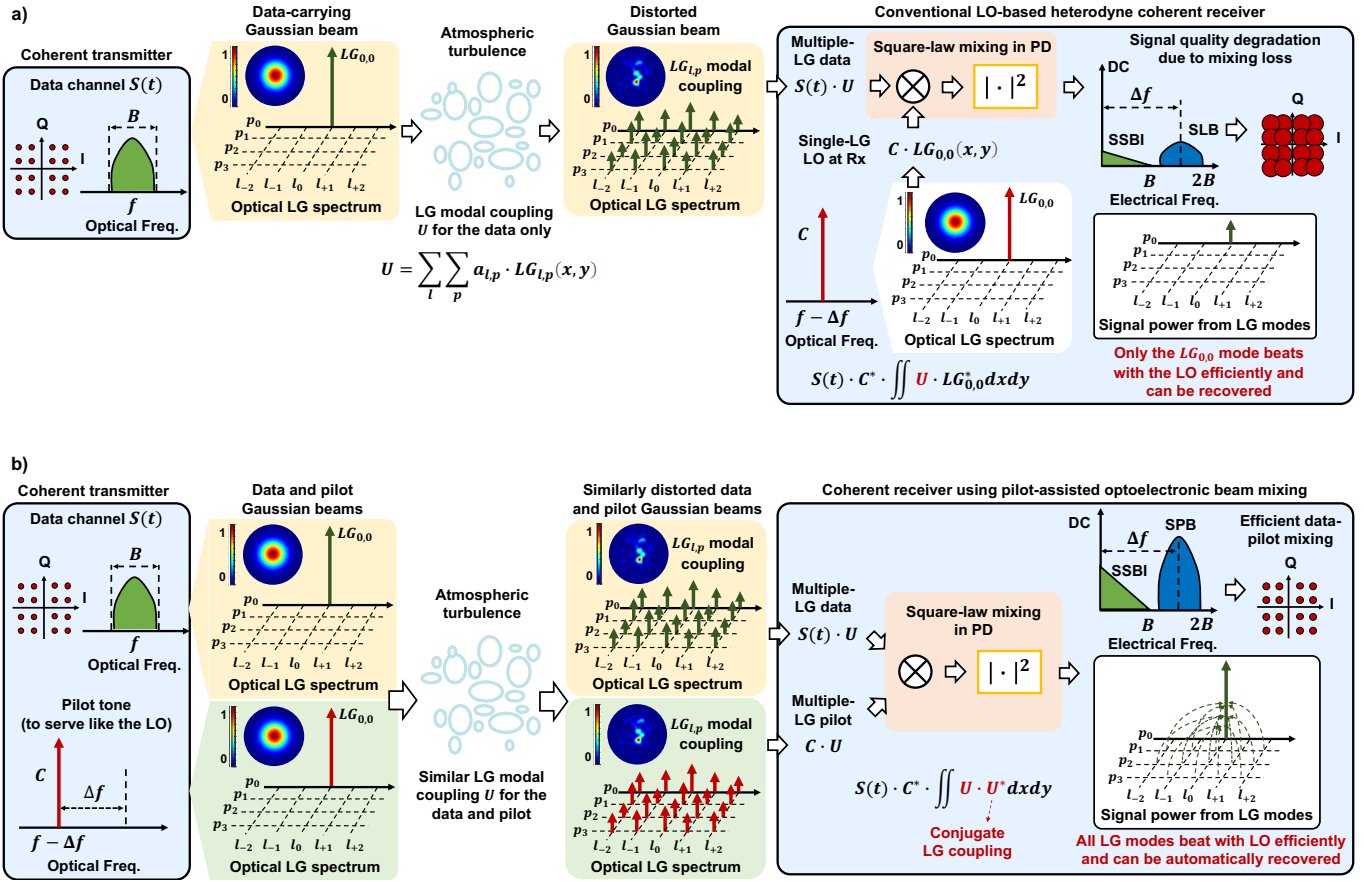

**Figure 1 | Concept of utilizing pilot-assisted optoelectronic beam mixing to automatically compensate the turbulence-induced LG modal power coupling in a coherent FSO link. (a)** A fundamental Gaussian beam (*i.e.*, $LG_{0,0}$ mode) carrying a 16-QAM data is transmitted through turbulent atmosphere. Due to the turbulence-induced LG modal power coupling, the received data beam would contain many LG modes. In a conventional LO-based heterodyne receiver, only the $LG_{0,0}$ mode can be efficiently mixed with the LO and recovered. **(b)** In the pilot-assisted optoelectronic beam mixing approach, we transmit additional CW pilot which experiences similar turbulence-induced LG coupling as the data beam and serves a similar function as the LO at the receiver. During mixing of the pilot and data beams in a square-law detector, a conjugate of the turbulence experienced by the pilot is automatically generated and compensates the turbulence experienced by the data beam. Therefore, all the data-carrying LG modes can be efficiently mixed with the pilot and recovered. DC: direct current; SSBI: signal-signal-beating interference; SLB: signal-LO beating; SPB: signal-pilot beating. In both cases, the frequency offset $\Delta f$ is greater than the data bandwidth $B$ to avoid the SSBI.

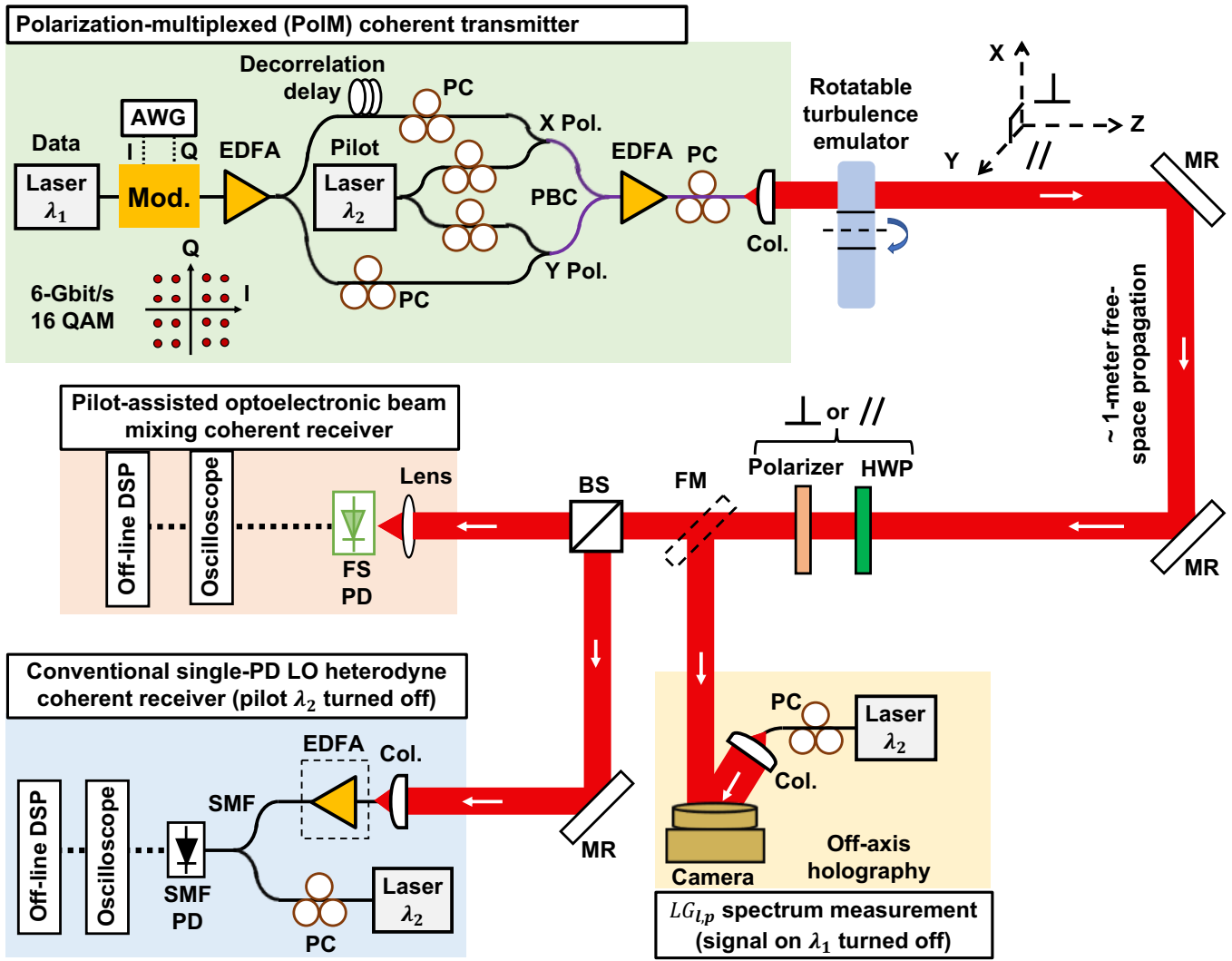

**Figure 2 | Experimental setup for 12-Gbit/s 16-QAM PDM coherent FSO link.** At the transmitter, pilot-data-channel pair is transmitted on each of the orthogonal polarizations. The PolM Gaussian beams then propagate through a rotatable turbulence emulator. At the receiver, the distorted beams are sent to (by a FM) an off-axis holography for LG spectrum measurement. To compare the performance of two different coherent receivers, equal copies of the received beams are detected by the pilot-assisted coherent detector and conventional single-PD LO heterodyne detector. During the detection of the conventional LO-based detector, the pilot is turned off. Same DSP algorithms are applied to both receivers to retrieve the 16-QAM data. AWG: arbitrary waveform generator; Mod.: modulator; EDFA: erbium-doped fiber amplifier; PC: polarization controller; PBC: polarization beam combiner; Col.: collimator; MR: mirror; HWP: half-wave plate; FM: flip mirror; FS PD: free-space-coupled photodetector; SMF: single-mode fiber; DSP: digital signal processing.

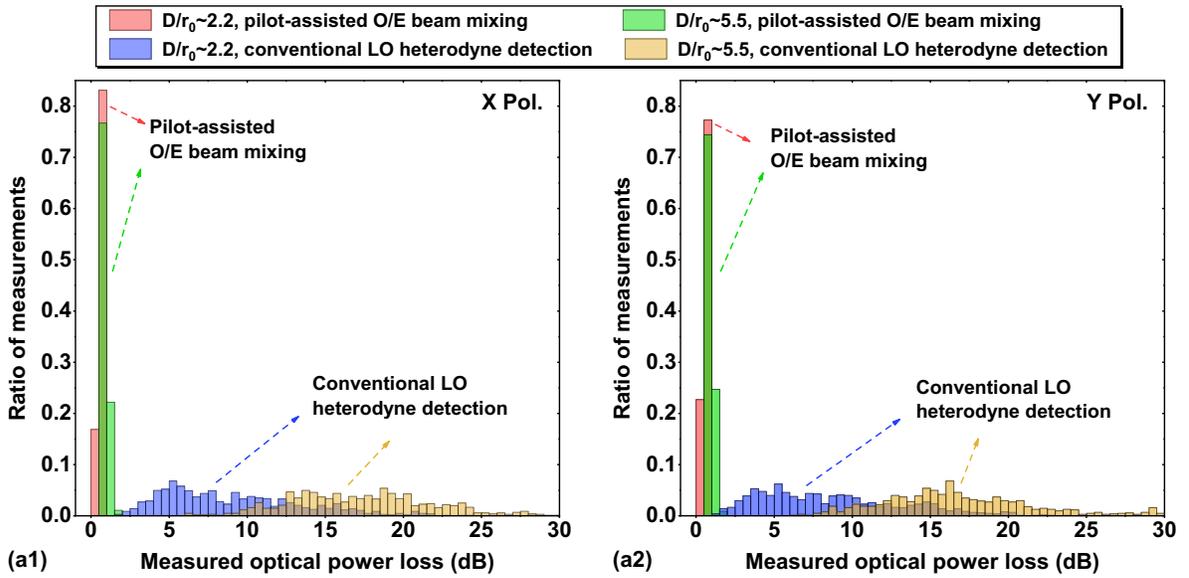

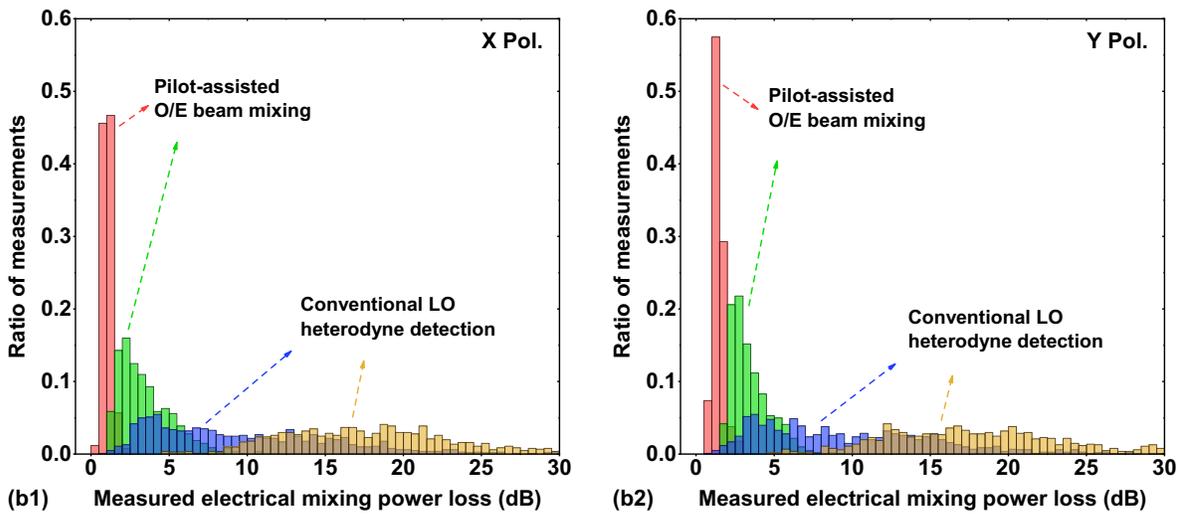

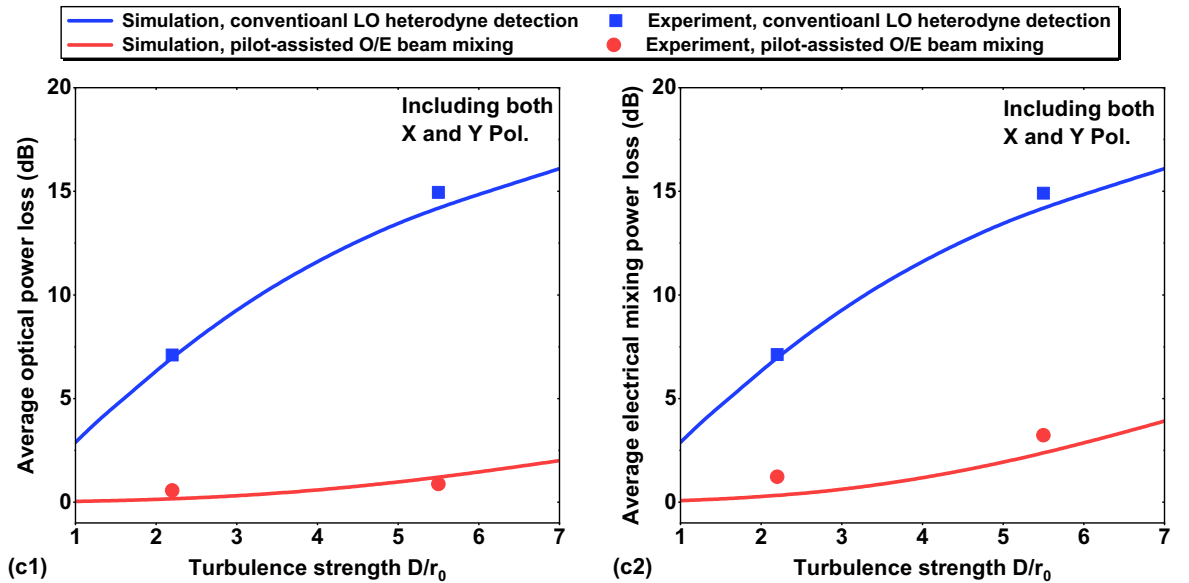

**Figure 3 | Characterization of optical and mixing power loss for a coherent FSO system under different turbulence strengths. (a)** Experimentally measured optical power loss under two different turbulence distortions ($D/r_0 \sim 2.2$ and $D/r_0 \sim 5.5$) using the pilot-assisted detector and conventional LO heterodyne detector for X and Y polarizations. **(b)** Experimentally measured mixing power loss (in the electrical domain) under two different turbulence distortions ($D/r_0 \sim 2.2$ and $D/r_0 \sim 5.5$) using the pilot-assisted detector and conventional LO heterodyne detectors for X and Y polarizations. The mixing power loss is measured at the IF of $\sim 2.6$ GHz in the electrical domain. We measure 1000 random turbulence realizations for each polarization in (a) and (b). **(c)** Simulated average power loss for different turbulence strength $D/r_0$ from 1 to 7. The average optical and mixing power loss are shown in (c1) and (c2), respectively. The average values of experimentally measured data points (including both X and Y polarizations) are also plotted as scatters in (c). O/E: optoelectronic.

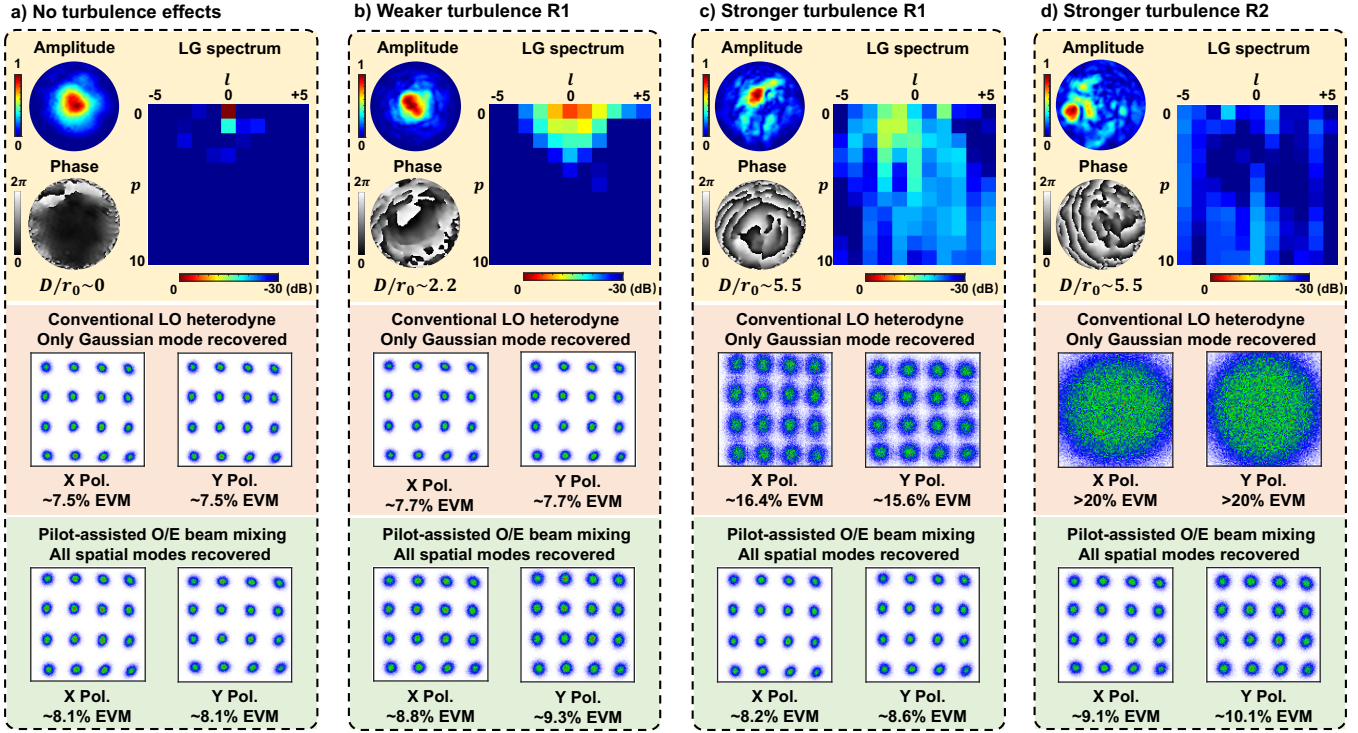

**Figure 4 | Experimental results of turbulence-induced LG modal power coupling and recovered 16-QAM data qualities using the conventional single-PD LO heterodyne detector and pilot-assisted coherent detector. (a)** No turbulence distortion; **(b)** One example realization of the weaker turbulence distortion ($D/r_0 \sim 2.2$); **(c)** and **(d)** Two different example realizations of the stronger turbulence distortions ($D/r_0 \sim 5.5$). For each of the four realizations, we measure the LG modal power spectrum (two indices $-5 \leq l \leq +5$ and $0 \leq p \leq 10$) and recover the 16-QAM data constellations. To compare the performance of the conventional LO-based and pilot-assisted detectors, we apply the same off-line DSP algorithms for both receivers. In this demonstration of PolM FSO data transmission, each polarization carries a 6-Gbit/s 16-QAM signal. R1 (R2): realization 1 (2); EVM: error vector magnitude. O/E: optoelectronic.

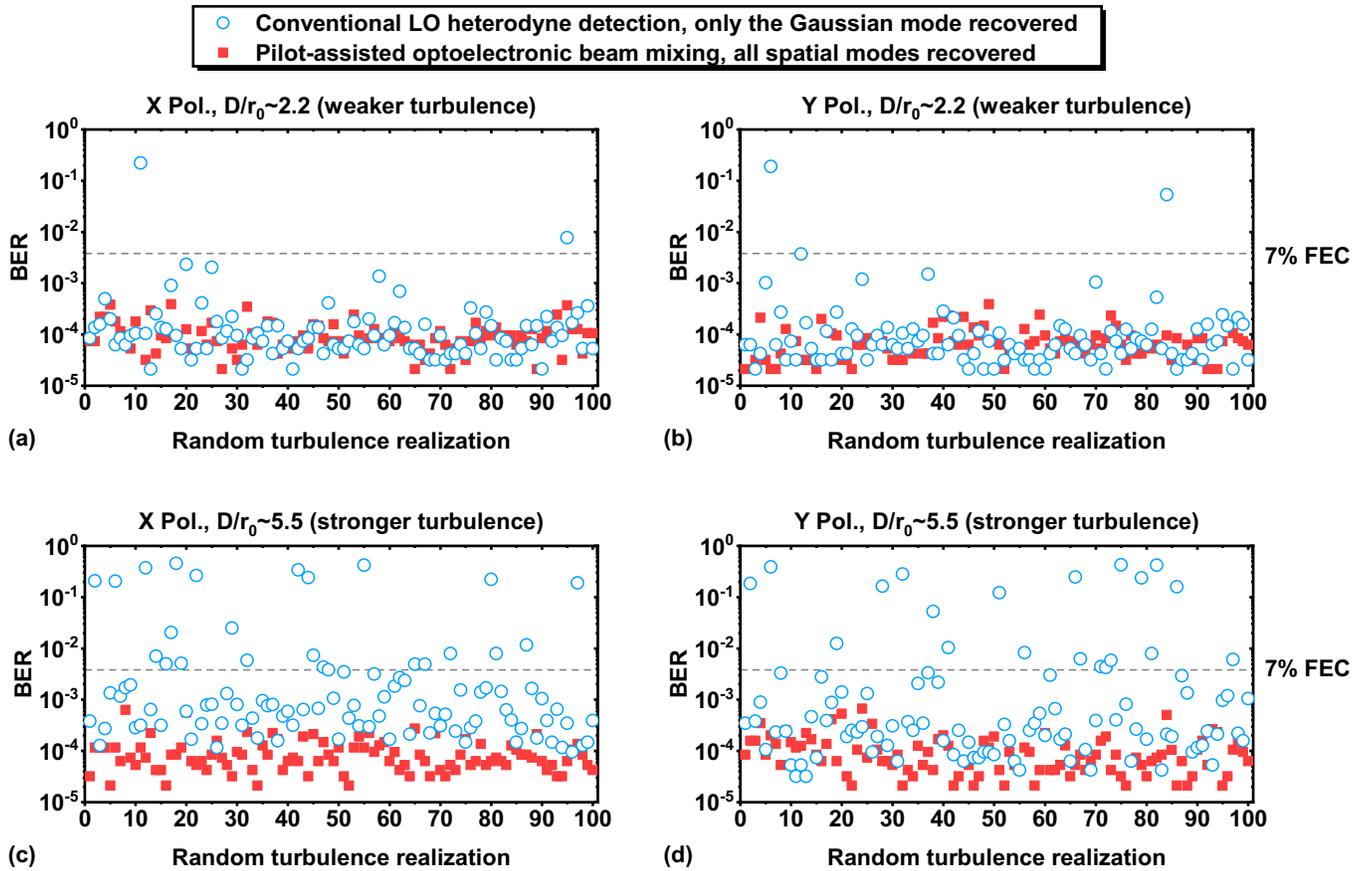

**Figure 5 | Experimentally measured BER performance of the 12-Gbit/s PolM FSO link over 200 different emulated random turbulence realizations.** **(a)** Weaker turbulence effects, X Pol; **(b)** Weaker turbulence effects, Y Pol; **(c)** Stronger turbulence effects, X Pol; **(d)** Stronger turbulence effects, Y Pol. The performance using the conventional single-PD LO heterodyne detector and the pilot-assisted coherent detector on each polarization are shown for comparison. The captured data signal is amplified by an EDFA at the receiver for the conventional LO heterodyne detector. The weaker and stronger turbulence strengths are $D/r_0 \sim 2.2$ and $D/r_0 \sim 5.5$, respectively. FEC: forward error correction.

# Supplementary Information: Turbulence-Resilient Coherent Free-Space Optical Communications using Automatic Power-Efficient Pilot-Assisted Optoelectronic Beam Mixing of Many Modes

**Runzhou Zhang[1]\*, Nanzhe Hu[1]\*, Huibin Zhou[1], Kaiheng Zou[1], Xinzhou Su[1], Yiyu Zhou[2], Haoqian Song[1], Kai Pang[1], Hao Song[1], Amir Minoofar[1], Zhe Zhao[1], Cong Liu[1], Karapet Manukyan[1], Ahmed Almaiman[1,3], Brittany Lynn[4], Robert W. Boyd[2], Moshe Tur[5], and Alan E. Willner[1]**

1. Department of Electrical and Computer Engineering, Univ. of Southern California, Los Angeles, CA 90089, USA
2. Institute of Optics, University of Rochester, Rochester, New York 14627, USA
3. King Saud University, Riyadh 11362, Saudi Arabia
4. Naval Information Warfare Center Pacific, San Diego, CA, 92152, USA
5. School of Electrical Engineering, Tel Aviv University, Ramat Aviv 69978, Israel
\* These authors contributed equally to this work.
Corresponding emails: R.Z. (runzhou@usc.edu) or A.E.W. (willner@usc.edu)

## Introduction

In this supplementary information, we present additional simulation and experimental results for the performance of the pilot-assisted coherent detector. In addition to the results shown in the main text, this supplementary information includes the following: (i) random-phase-screen (RPS) simulations for turbulence-induced optical power loss and electrical mixing power loss using the pilot-assisted coherent detector versus a conventional single-photodetector (PD) local-oscillator (LO) heterodyne detector; (ii) digital-image-processing procedures for extracting the complex wavefront of the pilot beam at the receiver aperture using off-axis holography; (iii) experimentally measured electrical spectra under example random turbulence realizations (same realizations in Fig. 4 of the main text); and (iv) experimentally measured bit-error-rate (BER) performance as a function of the transmitted optical power for the pilot-assisted coherent detector under no turbulence effects and one example random realization of the stronger turbulence effects (strength $D/r_0 \sim 5.5$).

The supplementary simulation results indicate the following when comparing the pilot-assisted coherent detector to the conventional single-PD LO heterodyne detector using the 1-RPS model: (a) the pilot-assisted detector has an optical power loss of up to ~3 dB while the conventional LO-based detector can suffer from an optical power loss of up to ~35 dB under 99% random realizations of the stronger turbulence (strength $D/r_0 \sim 5.5$); (b) the pilot-assisted detector has an electrical mixing power loss of up to ~6 dB while the conventional LO-based detector can suffer from a mixing power loss of up to ~35 dB under 99% random realizations of the stronger turbulence (strength $D/r_0 \sim 5.5$); and (c) these simulation results are generally in agreement with the experimental results; the simulated optical and electrical mixing power loss using the 5-RPS (*i.e.*, multiple phase screens) models exhibit similar trends and distributions as the 1-RPS simulation.

Furthermore, the supplementary experimental results for the measured electrical spectra indicate the following, compared to the case of no turbulence effects: (a) the conventional LO heterodyne detector can suffer from an electrical signal-to-noise-ratio (SNR) degradation of up to ~18 dB for both polarizations under one example random realization of the stronger turbulence effects (strength $D/r_0 \sim 5.5$); and (b) the pilot-assisted coherent detector can suffer from an electrical SNR degradation of up to ~3 dB for both polarizations under the same turbulence realization with a strength of 5.5. Moreover, the measured BER performance of the pilot-assisted coherent detector shows that: (a) it can achieve below the 7% forward-error-correction (FEC) limit under the cases of no turbulence and one example random realization of the stronger turbulence (strength $D/r_0 \sim 5.5$); and (b) compared to the case of no turbulence, we measure power penalties of ~3 dB for both X and Y polarizations under the stronger turbulence realization, each polarization carrying an independent 6-Gbit/s 16-quadrature-amplitude-modulation (QAM) data channel.

**RPS simulation for a fundamental Gaussian beam's propagation through atmospheric turbulence**

In the Results section of the main text, we show the simulated average optical and electrical mixing power loss for both the conventional LO-based and pilot-assisted detectors under two different turbulence strengths ($D/r_0 \sim 2.2$ and 5.5) in Fig. 3. Here we present additional simulation results for the probability distribution of optical and mixing power loss for both detectors induced by the tubulence effects using the single- or multiple-RPS model separately[1].

In general, atmospheric turbulence effects can be simulated using split-step RPS models by either: (i) a single phase screen for the entire link[1] or (ii) multiple phase screens with each screen representing one portion of the link[1,2]. In general, larger number of the phase screens tends to increase the accuracy of the turbulence emulation[1,2]. To indicate the relative accuracy of our turbulence emulation method, we simulate: (a) the 1-RPS turbulence effects to validate our experimental measurements, and (b) the 5-RPS turbulence effects to show similar trends for the turbulence-induced optical and mixing power loss as the 1-RPS simulation. In our simulation, the phase distribution of each phase screen follows Kolmogorov power spectrum statistics, and the beam propagation is simulated using Fresnel diffraction[1]. For each turbulence strength $D/r_0$, we simulate 10,000 random turbulence realizations to calculate the probability distributions and average values of optical power loss and electrical mixing loss.

We numerically find the optical power loss and the electrical mixing power loss for both the conventional single-PD LO heterodyne detector and the pilot-assisted coherent detector. As mentioned in the Introduction section of the main text, the conventional LO-based detector can either be coupled to: (i) single-mode fiber (SMF), in which case the higher-order modes are not efficiently captured by the detector[3], or (ii) free-space, in which case the higher-order modes are detected but don't efficiently mix with the LO[4]. For simplicity, we model below the fiber-coupled conventional LO-based detector, although comparative results and trends for the free-space-coupled case would be in general agreement.

The Laguerre-Gaussian (LG) optical fields ($LG_{l,p}$) used in the simulation are expressed in Suppl. Eq. (1)[5]:

$$LG_{l,p} = \frac{A_{lp}^{LG}}{w(z)}\left(\frac{r\sqrt{2}}{w(z)}\right)^{|l|}\exp\left(-\frac{r^2}{w^2(z)}\right)L_p^{|l|}\left(\frac{2r^2}{w^2(z)}\right)$$
$$\exp\left(-ik\frac{r^2}{2R(z)}\right)\exp(-il\varphi)\,exp\,(-ikz)exp\,(i\psi(z)),$$

(1)

where $A_{lp}^{LG}, L_p^l, R(z), w(z)$ and $\psi(z)$ are the normalization constant, generalized Laguerre polynomials, curvature radius, beam width, and Gouy phase, respectively.

For the conventional single-PD LO-based detector, we calculate the optical power loss using the following Suppl. Eq. (2)[3]:

$$Optical\ power_{\text{conventional}} \propto P_{\text{data in SMF}}$$
$$\propto \left|\iint E_{\text{data}}(x,y)\cdot Aperture(x,y)\cdot LG_{0,0}^*(x,y)dxdy\right|^2,$$

(2)

where $P_{\text{data in SMF}}$ is the optical power coupled to the SMF-based PD; $E_{\text{data}}(x,y)$ and $LG_{0,0}(x,y)$ are the simulated complex field of the distorted data beam and the theoretical complex field of the $LG_{0,0}$ mode, respectively; $Aperture(x,y)$ is the aperture function of the receiver. For the pilot-assisted coherent detector, we calculate the optical power loss for the pilot-assisted detector using the following Suppl. Eq. (3):

$$Optical\ power_{\text{pilot-assisted}} \propto P_{\text{data at Rx}}$$
$$\propto \iint|E_{\text{data}}(x,y)|^2\cdot Aperture(x,y)dxdy,$$

(3)

where $P_{\text{data at Rx}}$ is the optical power coupled to the receiver's aperture.

The following equations address the mixing power in electrical domain. For the conventional single-PD LO-based heterodyne detector, we calculate the mixing power loss using the Suppl. Eq. (4)[6]:

$$Mixing\ power_{\text{conventional}} \propto P_{\text{data in SMF}}\cdot P_{\text{LO}}$$
$$\propto \left|\iint E_{\text{data}}(x,y)\cdot Aperture(x,y)\cdot LG_{0,0}^*(x,y)dxdy\right|^2\cdot\left|\iint E_{\text{LO}}(x,y)dxdy\right|^2,$$

(4)

where the $P_{\text{data in SMF}}$ and $P_{\text{LO}}$ represent the optical power of the data beam and LO beam coupled to the SMF-coupled PD, respectively; $E_{\text{LO}}(x,y)$ is the complex field of the LO beam. For the pilot-assisted coherent detector, we calculate the mixing power loss using the Suppl. Eq. (5)[7,8]:

$$Mixing\ power_{\text{pilot-assisted}} \propto P_{\text{data at Rx}} \cdot P_{\text{pilot at Rx}} \cdot \eta, \tag{5}$$

where $P_{\text{data at Rx}}$ and $P_{\text{pilot at Rx}}$ are the received optical power of the data beam and pilot beam, respectively; $\eta$ is the mixing efficiency, as defined in Suppl. Eq. (6)[7,8]:

$$\eta = \frac{\left| \iint E_{\text{data}}(x,y) \cdot E_{\text{pilot}}^*(x,y)dxdy \right|^2}{\left| \iint E_{\text{data}}(x,y)dxdy \right|^2 \cdot \left| \iint E_{\text{pilot}}(x,y)dxdy \right|^2}, \tag{6}$$

where $E_{\text{data}}$ and $E_{\text{pilot}}$ are the simulated optical fields of the received data beam and pilot beam, respectively. Note that we assume the optical power of the LO, pilot, and data beams to be fixed in this simulation.

Supplementary Fig. 1 shows the simulation results for optical power loss using the conventional LO-based and pilot-assisted detectors for turbulence strengths $D/r_0 \sim 1.0$, 2.2, and 5.5. As shown in Suppl. Fig. 1(a1) and (b1), the simulated turbulence-induced optical power loss exhibits similar trends and power distributions as the experimentally measured power values in Fig. 3(a) of the main text. Note that the scales of horizontal and vertical axises are different in Suppl. Fig. 1(a) and (b). As stated in the main text, the turbulence effects can induce power coupling from the fundamental Gaussian mode to many higher-order LG modes[3]. Since the LO-based detector recovers only the Gaussian mode, it can suffer for some realizations a large optical power loss, *e.g.*, a loss of up to ~25.5 dB and ~35.6 dB for 90% and 99% random realizations when the turbulence strength is $D/r_0 \sim 5.5$. However, the pilot-assisted detector has an optical power loss only of up to ~1.8 dB and ~ 3.0 dB for 90% and 99% random realizations with the same turbulence strength of ~5.5, as shown in Suppl. Fig. 1(b1). This is because the free-space-coupled PD of the modeled pilot-assisted approach can capture almost the entire distorted beam profile as compared to the SMF-based PD of the modeled conventional LO-based approach. Moreover, as shown in Suppl. Fig. 1(a2), (b2), and (c), we find generally similar system degradation trends for both detectors when comparing the models of 1-RPS to 5-RPS cases.

For the turbulence strength $D/r_0 \sim 5.5$, we find a higher average optical power loss of ~2.7 dB and a lower average optical power loss of ~0.9 dB for the conventional LO-based and pilot-assisted detectors, respectively, when comparing the 5-RPS simulation to the 1-RPS simulation.

Supplementary Fig. 2 illustrates the simulated electrical mixing power loss using the conventional LO-based heterodyne detector and pilot-assisted coherent detector. As shown in Suppl. Fig. 2(a1) and (b1), the simulated turbulence-induced electrical mixing power loss exhibits similar trends and distributions as the experimentally measured values in Fig. 3(b) of the main text. Note that the scales of horizontal and vertical axises are different in Suppl. Fig. 2(a) and (b). As shown in Suppl. Fig. 2(a1), the conventional LO heterodyne detector can have a mixing power loss of up to ~25.5 dB and ~35.6 dB for 90% and 99% random realizations when the turbulence strength is $D/r_0 \sim 5.5$. However, the pilot-assisted detector has a mixing power loss only of up to ~3.6 dB and ~6.1 dB for 90% and 99% random realizations with the same turbulence strength of ~5.5, as shown in Suppl. Fig. 2(b1). This is due to that the pilot-assisted detector can efficiently mix and recover almost all the captured data modes. Moreover, as shown in Suppl. Fig. 2(a2), (b2), and (c), we find generally similar trends and distributions of mixing power loss for both detectors using the 1-RPS and 5-RPS simulations. For the turbulence strength $D/r_0 \sim 5.5$, we find a higher average electrical mixing power loss of ~2.7 dB and a lower average mixing power loss of ~1.6 dB for the conventional LO-based and pilot-assisted detectors, respectively, when comparing the 5-RPS simulation to the 1-RPS simulation.

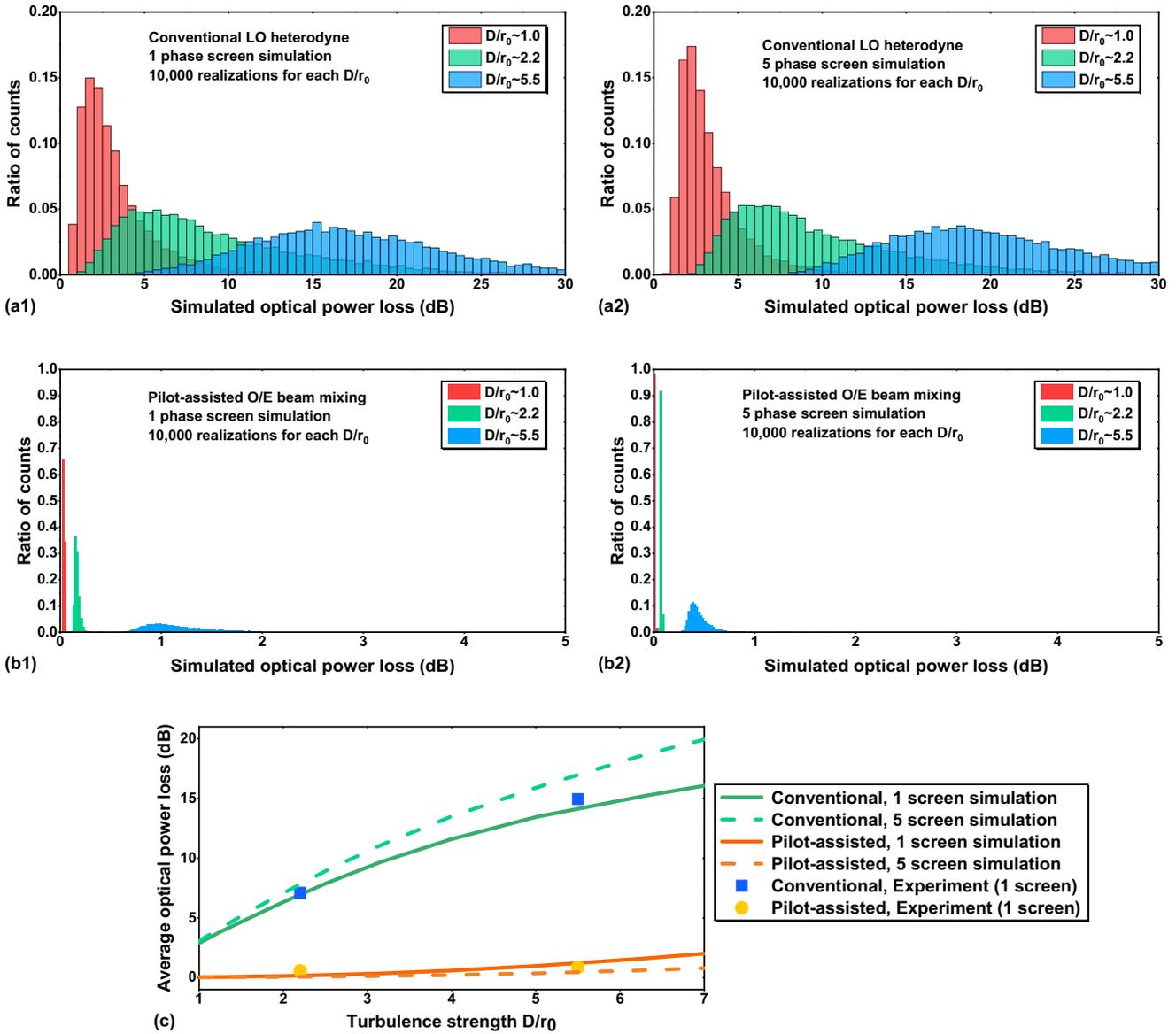

**Supplementary Figure 1 | Simulation results of optical power loss using random-phase-screen model.** (**a**) Optical power loss using the conventional LO-based heterodyne detector. 1 and 5 random phase screens are simulated in (a1) and (a2), respectively. (**b**) Optical power loss using the pilot-assisted detector. 1 and 5 random phase screens are simulated in (b1) and (b2), respectively. The scales of horizontal and vertical axes are different in (a) and (b). (**c**) Average optical power loss using the conventional LO-based and pilot-assisted detectors. The average value for each $D/r_0$ is calculated over 10,000 random realizations. The experimentally measured values of the average optical power loss using the turbulence emulator (1 phase screen) are also plotted as scatters in (c). O/E: optoelectronic.

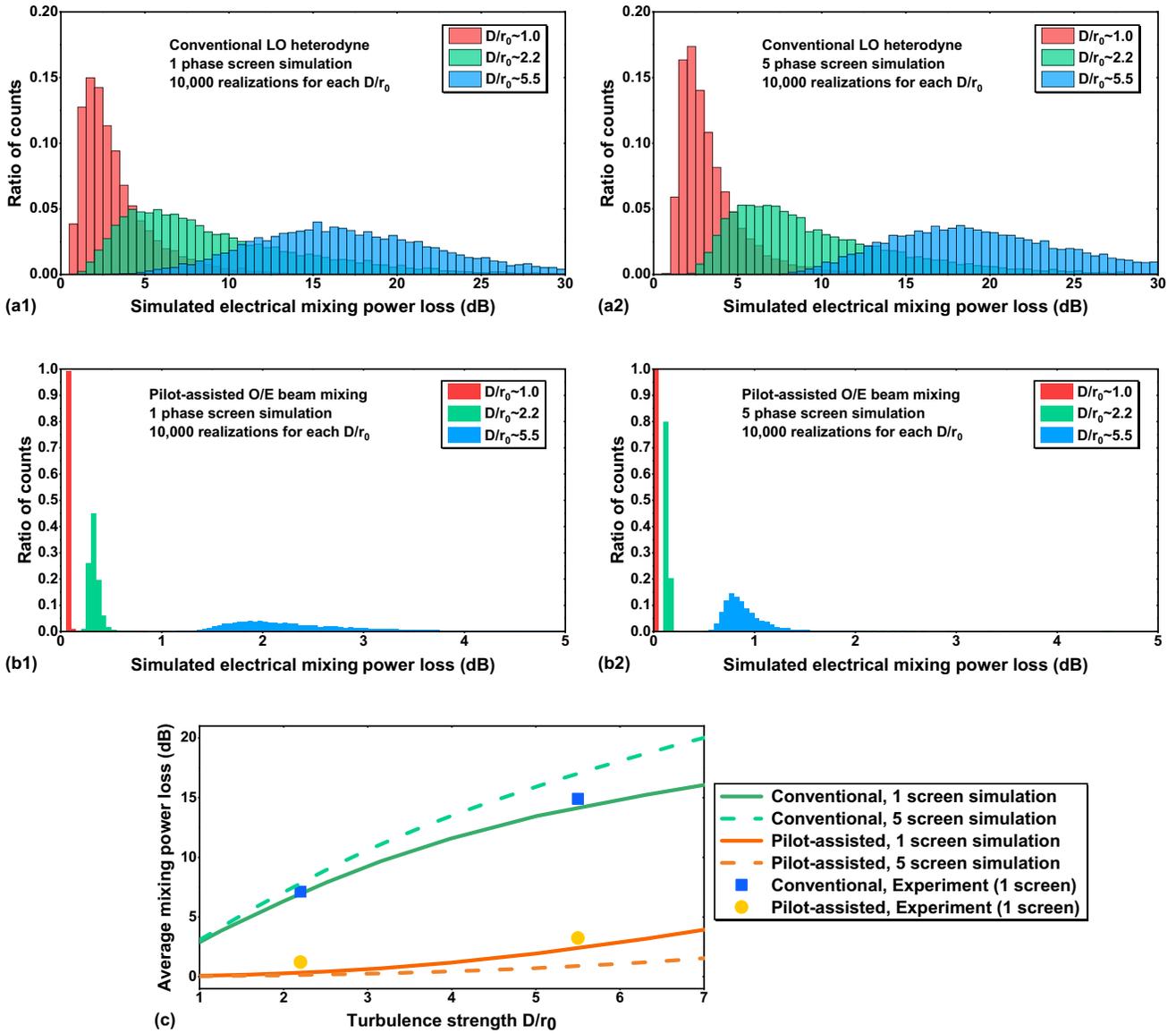

**Supplementary Figure 2 | Simulation results of electrical mixing power loss using random-phase-screen model.** (**a**) Mixing power loss using the conventional LO-based heterodyne detector. 1 and 5 random phase screens are simulated in (a1) and (a2), respectively. (**b**) Mixing power loss using the pilot-assisted detector. 1 and 5 random phase screens are simulated in (b1) and (b2), respectively. The scales of horizontal and vertical axises are different in (a) and (b). (**c**) Avergare mixing power loss using the conventional LO-based and pilot-assisted detectors. The average value for each $D/r_0$ is calculated over 10,000 random realizations. The experimentally measured values of the average mixing power loss using the turbulence emulator (1 phase screen) are also plotted as scatters in (c). O/E: optoelectronic.

**Digital-image-processing procedures for measuring the LG spectrum of the turbulence-distorted beam using off-axis holography**

In the Results section of the main text, we show the measured LG spectra of the distorted pilot beam under different turbulence realizations. Here we present the step-by-step digital-image-processing procedures we use to extract the complex wavefront of a distorted beam using the off-axis holography.

As shown in Suppl. Fig. 3, we perform the following steps to extract both the spatial amplitude and phase profiles of the distorted pilot beam[9]:

(i) Record the interferogram between the distorted pilot beam and another off-axis undistorted reference Gaussian beam using an infrared camera.

(ii) Perform two-dimensional Fourier transform of the interferogram to obtain the spatial frequency spectrum, filter out the 1st-order diffraction, and shift the 1st-order diffraction to the center of the spatial frequency spectrum.

(iii) Perform two-dimensional inverse Fourier transform of the shifted spatial frequency spectrum, and subsequently obtain the spatial amplitude and phase profiles of the distorted beam (*i.e.*, $E_{\mathrm{rec}}(x, y)$).

(iv) Decompose the distorted pilot beam in the two-dimensional LG modal basis using the Suppl. Eq. (7)[10]:

$$a_{l,p} = \iint E_{\mathrm{rec}}(x, y) \cdot LG_{l,p}^*(x, y) dx dy,　\tag{7}$$

where $E_{rec}(x, y)$ and $LG_{l,p}(x, y)$ are the measured complex field of the distorted pilot beam and the theoretical complex field of an $LG_{l,p}$ mode, respectively. The ratio of optical power coupling to the $LG_{l,p}$ mode is given by $\left| a_{l,p} \right|^2$.

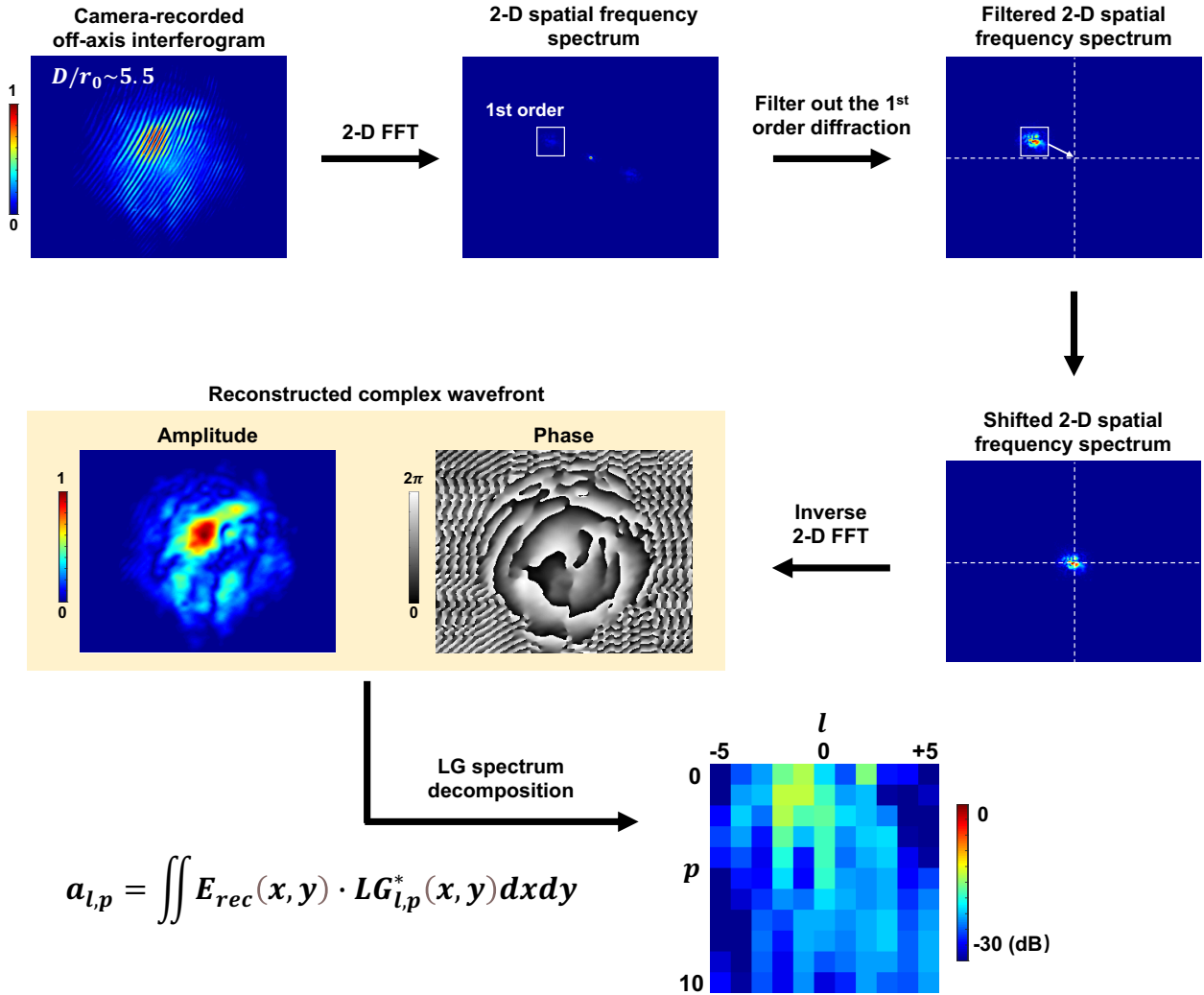

**Supplementary Figure 3 | The procedures for measuring the LG spectrum of a distorted pilot beam using the off-axis holography.** The spatial amplitude and phase profiles of the distorted pilot beam are obtained, and subsequently the corresponding LG power spectrum is calculated. The captured images have $320 \times 256$ pixels, with a pixel size of 30 μm. FFT: fast Fourier transform; $a_{l,p}$: coupling coefficient for the $LG_{l,p}$ mode. The ratio of optical power coupled to the $LG_{l,p}$ mode is given by $\left| a_{l,p} \right|^2$.

**Measured electrical spectra for the conventional single-PD LO heterodyne detector and pilot-assisted coherent detector under example random realizations of the weaker and stronger turbulence**

In the Results section of the main text, we show the recovered data channel's constellation diagrams using the conventional LO heterodyne detector and pilot-assisted detector under example realizations of the weaker and stronger turbulence in Fig. 4. To further illustrate the turbulence-resilient performance of the pilot-assisted coherent detector, here we present the correspondingly measured electrical spectra (including the signal-LO-beating (SLB) or the signal-pilot-beating (SPB) photocurrent) for these example turbulence realizations in Suppl. Fig. 4. The turbulence realizations (and the measured LG spectra) shown in this figure correspond to the same realizations in Fig. 4 of the main text. Note that the optical signal is amplified by an erbium-doped fiber amplifier (EDFA) before detected by the PD (fixed optical power for data beam ~0 dBm) in the conventional LO heterodyne receiver.

As shown in Suppl. Fig. 4, for both conventional LO-based and pilot-assisted detectors, the electrical spectra include the signal-signal-beating-interference (SSBI) photocurrent located at from 0 to ~1.6 GHz and the desired SLB or SPB photocurrent located at from ~1.8 GHz to ~3.5 GHz. The SSBI and SLB (SPB) components indicate the amount of electrical data power that is detected by the PD and that can be utilized for data recover by the PD, respectivey. Without turbulence effects (Suppl. Fig. 4(a)), both detectors can achive efficient mixing of the data and LO (pilot) beams, resulting in suffient SLB (SPB) IF signal to recover the data channel. We note that the non-flat electrical spectra of the SPB signal for the pilot-assisted coherent detector is due to the non-flat frequency response of the free-space-coupled PD used in this experiment.

Under one example realization of the weaker turbulence ($D/r_0 \sim 2.2$, in Suppl. Fig. 4(b)), the power of the data beam is mainly coupled to some neighboring LG modes, the SLB power of the conventional LO detector is not significantly degraded because the received optical signal is amplified by an EDFA before being sent to the mixing in PD. However, under two example realizations of the stronger turbulence ($D/r_0 \sim 5.5$, in Suppl. Fig. 4(c) and Suppl. Fig. 4(d)), the turbulence effects can induce >25 dB optical power loss for the data beam (as indicated by the low SSBI power) and subsequenttly degrade the SNR of the desired SLB IF signal in the electrical domain. Specifically, as shown in Suppl. Fig. 4(d), little mixing power at the IF frequency (with an increased noise power level) can be utilized by the conventional LO-based detector and the data channel's information can hardly be recovered. Under this realization, the conventional LO detector can suffer from an SNR degradation of ~18 dB in the electrical domain for both polariztions, compared to the case of no turbulence effects.

With respect to the pilot-assisted coherent detector, although turbulence effects can induce power coupling to a large number of LG modes, the pilot-assisted detector can efficiently mix and recover all the data modes. Therefore, the mixing power of the SPB IF signal in the electrical domain exhibits resilience to all these example turbulence realizations. Moreover, since the detector can capture almost the entire distorted beam, the SSBI power is also not significantly affected by different turbulence realizations. For these three realizations shown in Suppl. Fig. 4(b), 4(c), and 4(d), due to efficient mixing of the data and pilot beams, sufficient SPB power is available for the pilot-assisted detector to recover the data channel with recovered data quality not strongly dependent on turbulence strength. Specifically, the pilot-assisted detector can have an SNR degradation of up to ~3 dB in the electrical domain for both polarizations under these example random realizations, compared to the case of no turbulence effects.

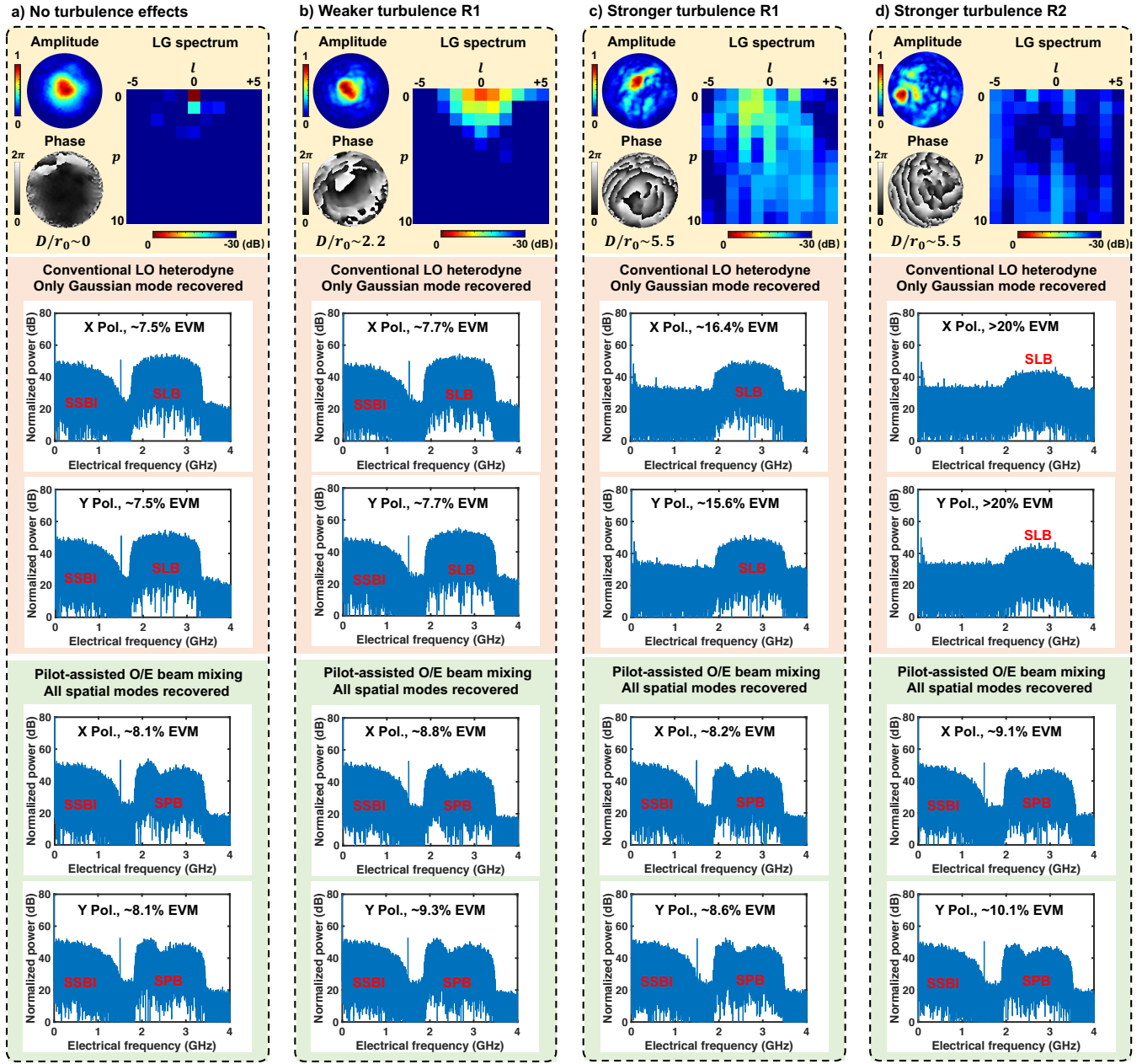

**Supplementary Figure 4 | Measured electrical spectra for the conventional single-PD LO heterodyne detector and the pilot-assisted coherent detector under example turbulence realizations (same realizations as shown in Fig. 4 of the main text). (a)** No turbulence distortion; **(b)** One example realization of the weaker turbulence distortion ($D/r_0 \sim 2.2$); **(c)** and **(d)** Two different example realizations of the stronger turbulence distortions ($D/r_0 \sim 5.5$). For the conventional LO heterodyne detector, the optical signal is amplified by an EDFA before mixed wih the LO in PD. SSBI: signal-signal-beating interference; SLB: signal-

LO-beating; SPB: signal-pilot-beating; EVM: error vector magnitude. The LG spectra shown in this figure are the same as the LG spectra shown in Fig. 4 of the main text.

**BER performance as a function of the transmitted optical power for the pilot-assisted coherent detector under stronger turbulence effects**

In the Results section of the main text, we show BER performance with a fixed transmitted optical power for the pilot-assisted coherent detector under different random turbulence realizations in Fig. 5. To further investigate its system performance, we measure the BER performance as a function of the transmitted optical power for the pilot-assisted coherent detector. In these measurements, we transmit 6-Gbit/s 16-QAM data on each of the two orthogonal polarizations, resulting in a 12-Gbit/s polarization-multiplexed link.

Supplementary Fig. 5 illustrates the measured BER performance for the pilot-assisted coherent detector without the turbulence effects and under one example realization of the stronger turbulence ($D/r_0 \sim 5.5$). Note that the transmitted optical power is the total power of the data and pilot beams at one polarization. For both X and Y polarizations, the pilot-assisted detector can achieve below the 7% FEC limit under the stronger turbulence realization example. Compared to the case of no turbulence effects, the BER performance for the pilot-assisted detector shows power penalties of $\sim 3$ dB at the 7% FEC limit (*i.e.*, 3.8e-3) for X and Y polarizations.

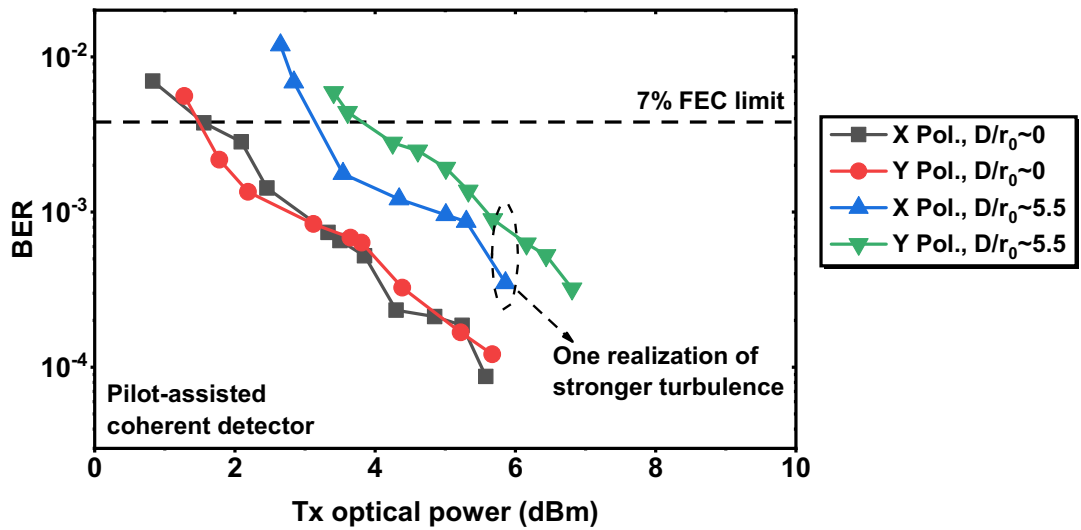

**Supplementary Figure 5 | BER as a function of the transmitted optical power for the pilot-assisted coherent detector.** The BER performance under the cases of no turbulence distortion ($D/r_0 \sim 0$) and one example realization of the stronger turbulence ($D/r_0 \sim 5.5$) are shown in the figure. Each polarization carries an independent 6-Gbit/s 16-QAM data channel. Tx: transmitter; FEC: forward error correction.